\documentclass[prl,superscriptaddress,floatfix,amsmath,notitlepage,twocolumn]{revtex4-1}
\usepackage{graphicx}
\usepackage{color}
\usepackage{siunitx}
\usepackage[normal]{subfigure}
\usepackage[usenames,dvipsnames]{xcolor}
\usepackage{amssymb}
\newcommand{\be}{\begin{equation}}
\newcommand{\ee}{\end{equation}}

\usepackage[colorlinks=true,urlcolor=blue,linkcolor=blue,citecolor=blue]{hyperref}

\begin{document}

\title{Quantum networks with chiral light--matter interaction in waveguides}

\author{Sahand Mahmoodian}
\author{Peter Lodahl}
\author{Anders~S.~S{\o}rensen}
\affiliation{Niels Bohr Institute, University of Copenhagen, Blegdamsvej 17, DK-2100 Copenhagen, Denmark}

\date{\today}

\begin{abstract}
We propose a scalable architecture for a quantum network based on a simple on-chip photonic circuit that performs loss-tolerant two-qubit measurements. The circuit consists of two quantum emitters positioned in the arms of an on-chip Mach-Zehnder interferometer composed of waveguides with chiral light--matter interfaces. The efficient chiral light--matter interaction allows the emitters to perform high-fidelity intranode two-qubit parity measurements within a single chip, and to emit photons to generate internode entanglement, without any need for reconfiguration. We show that by connecting multiple circuits of this kind into a quantum network, it is possible to perform universal quantum computation with heralded two-qubit gate fidelities ${\cal F} \sim 0.998$ achievable in state-of-the-art quantum dot systems.
\end{abstract}

\maketitle

The overarching goal of modern quantum physics is to construct large-scale quantum networks \cite{Kimble2008QuantumInternet} where the evolution and interaction of its constituents can be controlled at the quantum level. Such systems would provide a platform to simulate arbitrary quantum systems \cite{Feynman1982IJTP}, perform quantum computation \cite{Ladd2010Nature}, and provide provably secure information transfer \cite{GisinRMP2002}. Building a monolithic quantum system with a large number of qubits is a very challenging task, while controlled quantum operations on a small number of qubits with high fidelity is a feasible objective. An attractive approach to scale to a large quantum system is thus to construct a large number of few-qubit nodes and connect them with optical links \cite{Sorensen1998PRA, Cirac1999PRA, Jiang2007PRA, Monroe2014PRA}. In such systems it has been shown that quantum processing can successfully proceed even if the optical links are noisy \cite{Cirac1999PRA, Jiang2007PRA}. Building a scalable network of such few-qubit nodes becomes possible if one can combine an efficient light--matter interface with high-fidelity quantum operations between qubits within a single node.

In recent years quantum nanophotonic systems have matured greatly and photonic nanostructures have been interfaced efficiently with quantum dots  \cite{Lodahl2015RMP, Reithmaier2015NL}, atoms \cite{Goban2014NCOM, Junge2013PRL}, and diamond colour centres \cite{Hausmann2013NL}. In particular, photons in photonic-crystal waveguides (PCWs) can be interfaced with quantum dots (QDs) with near-unity efficiency \cite{Arcari2014PRL}, and can also exhibit chiral light--matter interaction where optical transitions with opposite helicity couple efficiently to opposite directions \cite{Sollner2015NNANO, Mitsch2014NCOM, Coles2016NCOM, Lobanov2015PRB}. A multitude of two-qubit gates have also been proposed for photon-emitter systems, however, there is no clear path for how to scale these ingredients to build large networks. In particular, since most schemes involve photon transfer between qubits to implement gates over long distances, it is unclear how to scale the systems without on-chip reconfiguration and/or lossy in/outcoupling from chips.

\begin{figure}[!t]
\includegraphics[width=\columnwidth]{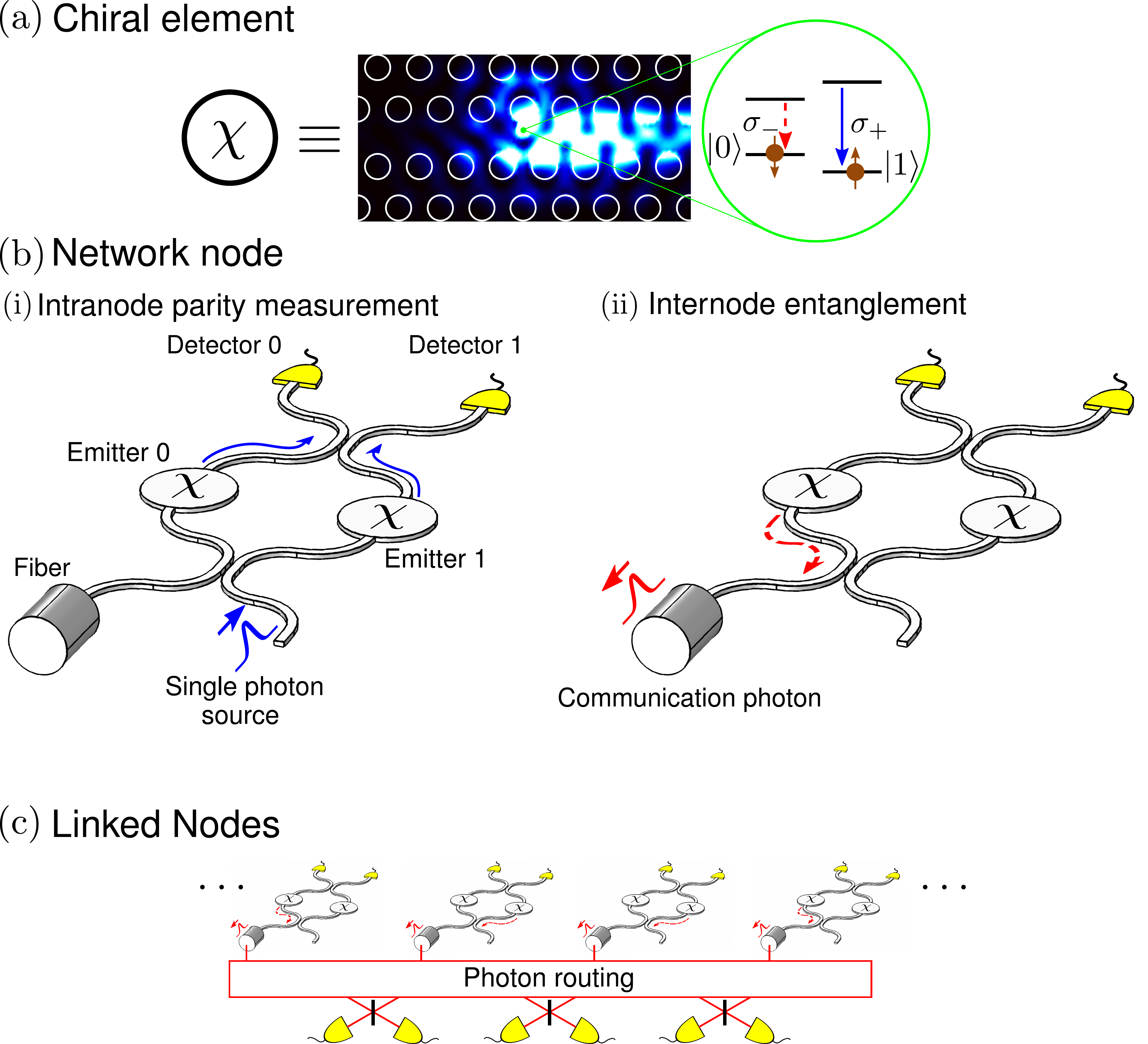}
\caption{\label{fig:chiralMZI}  Architecture for a quantum network. (a) The chiral element $\chi$ is composed of a quantum dot coupled to a photonic crystal waveguide. The quantum dot's optical transitions are circularly polarized with spin ground states that act as logical qubits and the waveguide is engineered to have chiral light--matter interaction \cite{Sollner2015NNANO}. A $\sigma_+$ ($\sigma_-$) polarized optical transition couples to the right (left) propagating waveguide mode. The electric field intensity emitted from a $\sigma_+$ dipole is shown with bright (dark) colours representing high (low) intensity. (b) The building block of the quantum network consists of two chiral elements $\chi$ in the arms of an interferometer with photon detectors at the output. (i) A two-qubit parity measurement within a node is performed using a single photon resonant on the $\sigma_+$ transition scattering off the emitters. The parity of the emitters' spin states determines the phase difference of the interferometer and thus the output path of the photon. (ii) Photon emission from the $\sigma_-$ transition is used to link nodes through an optical fiber. (c) Photons are routed to mix with the output from another node with photon detection heralding entanglement between any two nodes.}
\end{figure}

In this work, we propose an architecture for realizing a quantum network based on simple elements.  To this end we develop a novel scheme to measure the parity of two quantum dots' spin state in a manner that is robust to loss and achieve near-unity fidelity even for imperfect couplings. By linking many copies of these circuits together, we are able to realize a quantum network capable of universal quantum computation. The layout is shown in Figs.~\ref{fig:chiralMZI}(a)-(b) and consists of two QDs embedded in the arms of an on-chip Mach-Zehnder interferometer (MZI), with an additional single photon source. Here, dielectric waveguides couple to form beam splitters and are integrated with on-chip photon detectors. The waveguides are coupled to chiral elements $\chi$, Fig. \ref{fig:chiralMZI}(a), each of which contain a quantum emitter in a waveguide with chiral light--matter interaction. The modes of this waveguide have electric fields with in-plane circular polarization where counter-propagating modes have counter-circulating polarizations \cite{Sollner2015NNANO, Junge2013PRL}. In our proposal the chiral element is a QD coupled to an engineered PCW \cite{Sollner2015NNANO}, but other systems such as atoms coupled to whispering gallery mode resonators \cite{Junge2013PRL} can also be used. The logical qubits are formed by the spin ground states of the QDs, $|  \downarrow \rangle \equiv | 0 \rangle$ and $| \uparrow \rangle \equiv | 1 \rangle$, and the optical transitions are used both for gate operations and communication. Chiral light--matter interaction gives two important advantages: (i) it provides an efficient way to separate two optical transitions with opposite helicity, allowing each to be used for a distinct purpose. Here we use the right-hand circularly polarized ($\sigma_+$) transition for local operations and the left-hand circularly polarized ($\sigma_-$) transition for internode communication (Fig.~\ref{fig:chiralMZI}(c)). (ii) Our scheme requires a photon to coherently scatter off a quantum emitter and acquire a conditional phase shift before being passed on to the next element. The use of chiral interaction allows this to be achieved with near-unity fidelity in a simple passive structure without active switching. Similar functionality could in principle also be achieved with circulators or by exploiting the polarization in single-sided structures supporting circular polarization. Such elements would, however, be difficult to implement on chip.

%Similar schemes where photons scatter off a single-sided cavity or waveguide require bulk-optical elements to redirect the  photon to achieve near-unity efficiency \cite{Duan2004PRL, Witthaut2012EL}. Such elements are challenging to miniaturize and scale.  Here, the chiral light--matter interaction mediates non-reciprocal coupling \cite{Sollner2015NNANO, SayrinPRX2015} and such elements are not required.

Information is processed within a node using a single-photon source which sends a photon resonant with the $\sigma_+$ polarized transition into the MZI from the bottom-right arm (see Fig.~\ref{fig:chiralMZI}(b)(i)). This ancilla photon interacts with the two emitters and, as we detail below, measures the parity \cite{Beenakker2004PRL, Ionicioiu2007PRA} of the two qubits upon detection, i.e., it returns $0$ for  $| 0 0 \rangle$ and $| 1 1 \rangle$ and $1$ for $| 0 1 \rangle$ and $| 1 0 \rangle$, and projects the system into one of these subspaces. Since the operation is heralded by a photon detection, we show that losses act to lower the success probability of the measurement rather than its fidelity.  Finally, we show that such a parity measurement integrated in a simple passive circuit enables entanglement generation, swapping and distillation, as well as teleportation-based gates. These operations constitute a complete toolbox to perform universal quantum computation across a quantum network.

\begin{figure}[!t]
\includegraphics[width=\columnwidth]{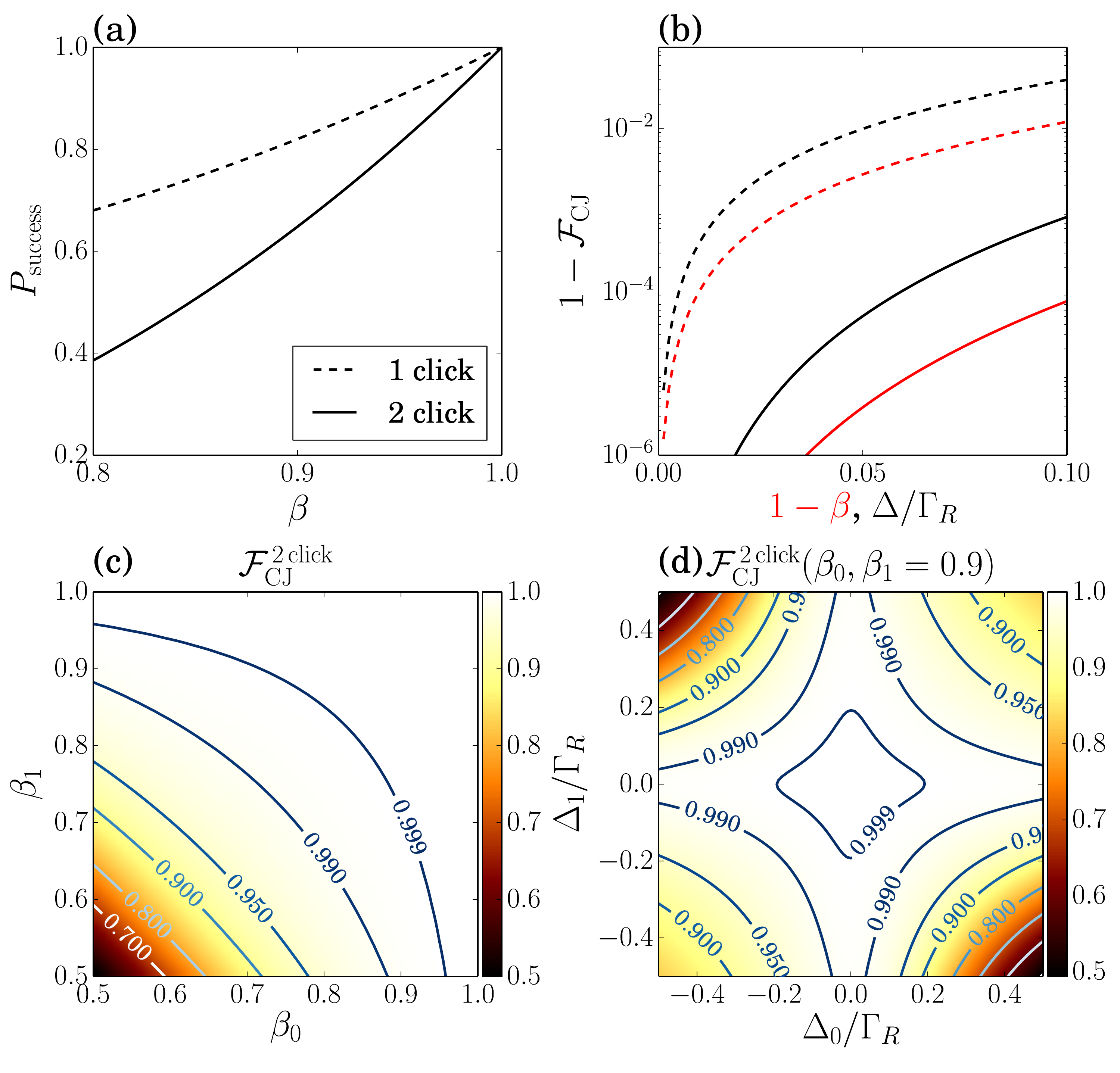}
\caption{\label{fig:gateAnalysis} Analysis of the parity measurement. (a) The success probability versus $\beta$-factor of the emitters ($\beta_0=\beta_1$) for the two protocols. (b)  Choi-Jamio{\l}kowski infidelity $1-{\cal F}_{\rm CJ}$ versus $1-\beta$ for resonant emitters with $\beta=\beta_0=\beta_1$  (light lines) and versus mutual detuning $\Delta = -\Delta_0 = \Delta_1$ (dark lines), for one-click (broken lines) and two-click (solid lines) protocols. Fidelity for two-click protocol ${\cal F}_{\rm CJ}^{\rm 2 click}$ versus,  (c) $\beta_0$ and $\beta_1$ for resonant emitters, and, (d) $\Delta_0/\Gamma_R$ and $\Delta_1/\Gamma_R$ with $\beta_0=\beta_1=0.9$.}
\end{figure}

We now consider the operation of the circuit in Fig.~\ref{fig:chiralMZI}(b)(i) for performing a parity measurement. The chiral coupling of the emitter to the waveguide is quantified by the directional $\beta$-factor of the $\sigma_+$ polarized optical transition to the rightward propagating waveguide mode $\beta = \Gamma_R/(\Gamma_R+\Gamma_L+\gamma)$, where $\Gamma_R$ and $\Gamma_L$ are the respective decay rates to the right and left-propagating waveguide modes and $\gamma$ is the decay rate to radiation modes. In the presence of a detuning $\Delta_j$, the transmission coefficient is $t_j = 1-2\beta_j/(1-2i\beta_j\Delta_j/\Gamma_R^j)$ \cite{Sollner2015NNANO, Fan2010PRA}, where the script $j$ denotes emitter 0 (left emitter in Fig.~\ref{fig:chiralMZI}(b)(i)) or 1 (right emitter). Ideally, narrowband on-resonance photons ($\beta_j=1$, $\Delta_j=0$) have unit transmittance and are imparted with a $\pi$ phase shift ($t_j=-1$) in each arm of the MZI if the emitter is in state $| 1 \rangle$. The ideal operation of the MZI for measuring parity is then clear: for odd-parity spin states ($| 0 1 \rangle$ and $| 1 0 \rangle$) the MZI is imbalanced and photons register clicks on Detector 1, while for even parity ($| 0 0 \rangle$ and $| 1 1 \rangle$) the MZI is balanced and clicks are registered on Detector 0. For an arbitrary input pure state $|\Psi\rangle = \sum_{i,j=0,1} c_{ij} |ij\rangle |1\rangle_{\rm ph}$, where $i$ and $j$ denote the states of the left and right stationary qubits respectively, the output state is
\begin{equation}
\begin{split}
\label{eq:out1}
&|\textrm{out} \rangle^{(1)} =  | 0 \rangle_{\rm ph} \left[ 2 c_{00}|00\rangle + (t_0+t_1) c_{11}|11\rangle \right]/2\\
&+ | 0 \rangle_{\rm ph} \left[(1+t_1) c_{01} |01\rangle + (1+t_0)c_{10} | 10 \rangle \right]/2\\
&+ | 1 \rangle_{\rm ph} \left[(1-t_0)c_{10} |10\rangle - (1-t_1) c_{01} |01\rangle \right]/2 \\
&+ | 1 \rangle_{\rm ph} \left[(t_0-t_1)c_{11}|11\rangle \right]/2 +|\textrm{lost}\rangle,
\end{split}
\end{equation}
where $| 0 \rangle_{\rm ph}$ ($| 1 \rangle_{\rm ph}$) represent a photon in the left (right) arm. Here, $|\textrm{lost} \rangle$ is not normalized and describes states where the photon is scattered into modes other than the right-propagating waveguide mode and is lost. The exact form of this term is not important as gate successes are heralded by detector clicks. The terms on the first and third lines of (\ref{eq:out1}) are the desired output states for a parity measurement, while the terms on the second and fourth lines of (\ref{eq:out1}) are erroneous results due to imperfections, i.e., the state of the ancilla photon is not equal to the parity of the qubits. The error terms on the second line occur due to imbalance in the MZI caused by loss ($\beta_0, \beta_1 < 1$) or detuning ($\Delta_0, \Delta_1 \neq 0$), while the error term on the fourth line occurs due to inequivalent emitters ($t_0 \neq t_1$).  For all errors, the probability of getting an erroneous click scales quadratically with its respective error parameter resulting in a very low probability, which scales as, e.g. $(1-\beta)^2$ and thus vanishes rapidly as $\beta \rightarrow 1$. The corresponding success probability and infidelity are shown in Figs.~\ref{fig:gateAnalysis}(a)-(b).

If even higher fidelity gate operation is required, errors caused by imbalances in the MZI can be minimized using a two-click protocol: the first photon detection is followed by a $\sigma_x$ operation (spin flip) on both qubits and a second photon is launched into the MZI. Two detection events on the same detector heralds success, otherwise the gate fails. The rotation and second photon scattering has three effects: (i) it reduces the probability of loss and detuning-induced errors to $\sim|1+t_0|^2|1+t_1|^2$. (ii) It removes the distortion in the states spanned by  $| 0 0 \rangle$ and $| 1 1 \rangle$ caused by different multiplicative factors of $|00\rangle$ and $|11 \rangle$ in the first line of (\ref{eq:out1}). (iii) It removes the error term due to inequivalent emitters. The reduced errors come at the expense of a reduced success probability with the two-click protocol requiring two successful detection events. A comparison of the success probabilities for one and two-click protocols is plotted versus the $\beta$-factor in Fig.~\ref{fig:gateAnalysis}(a). For detailed calculations, see the Supplementary Material (SM). We have also computed the fidelity of our implementation of the parity measurement. We compute the Choi-Jamiolkowski (CJ) fidelity which can be thought of as an average fidelity of all input states \cite{Gilchrist2005PRA, Das2015arXiv} (see SM for more information).  Figure~\ref{fig:gateAnalysis}(b) compares the CJ infidelity for one and two-click protocols versus $\beta$-factor and mutual detuning. The two-click protocol (solid lines) has significantly lower infidelity compared to the one-click protocol (dashed lines). Additionally, Fig.~\ref{fig:gateAnalysis}(c)-(d) shows the two-click protocol fidelity for emitters with different $\beta$-factors and detunings. In PCWs, coupling to non-guided radiation modes is suppressed and QDs can typically have $\beta \gtrsim 0.9$ \cite{Arcari2014PRL}, which for $\Delta=0$ gives an infidelity $1-{\cal F}_{\rm CJ}^{\rm 2 \, click} < 10^{-3}$ for the two click protocol. The fidelity is however more sensitive to detuning. Since the resonance frequencies of the QDs can be tuned electrically, we can expect $\Delta_1,\Delta_2<\Gamma_R/10$, which, for a $\beta$-factor $\beta=0.9$, gives ${\cal F}_{\rm CJ}^{\rm 2 \, click} \gtrsim 0.999$.

Until now we have only considered errors due to loss and detuning. Another source of error is incoherent scattering off the quantum emitter, which for QDs can be induced by pure dephasing and phonon-assisted scattering. These errors scale linearly with the fraction of incoherent scattering (see SM). Encouragingly, observations of photon emission with upto $99\%$ indistinguishability  \cite{Somaschi2015NPHOT, Matthiesen2013NCOM, He2013NNANO} and lifetime-limited linewidths \cite{Kuhlmann2013NPHYS} indicate that the effects of pure dephasing can be limited in QDs. However, phonon-assisted relaxation accounts for $\sim 10 \%$ of the emission in bulk QD samples \cite{Matthiesen2013NCOM, Konthasinghe2012PRB}. Operating the gate with high fidelity thus requires filtering out incoherently scattered photons. The spectral sideband associated with phonon assisted scattering has a maximum intensity that is detuned from the zero-phonon line by $\sim\SI{0.5}{\nano\metre}$ \cite{Konthasinghe2012PRB, Matthiesen2013NCOM}. The zero-phonon line can therefore be selectively filtered using a cavity with quality factor $Q\sim10^4$, which can be integrated on-chip \cite{MorichettiLPR2011} and positioned before the single photon detectors. Once the incoherent light is filtered out it only acts to lower the success probability and fidelity via a lower $\beta$-factor. Since these errors scale with the fourth power for the two-click protocol, the effect becomes small for phonon-assisted relaxation fractions $\lesssim 10\%$.

We now consider how the nodes in Fig.~\ref{fig:chiralMZI}(b) can be connected to perform quantum computation. The aim is to perform teleportation-based quantum computation based on nodes linked via entanglement \cite{Cirac1999PRA}. To this end, spontaneously emitted photons from the $\sigma_-$ polarized transition are used for inter-node communication since its emission couples to the left-propagating mode. This is shown schematically in Fig.~\ref{fig:chiralMZI}(b)-(c), where single photons emitted from adjacent nodes couple to output fibers and meet at a beamsplitter. Photon detection then heralds the generation of entanglement between the spin states in the two nodes \cite{Simon2003PRL}. Since the generation is heralded, it can be attempted until successful. We note that Beamsplitter 1 has a 50:50 splitting ratio so that half the photons are lost, but its efficiency can be improved by reconfiguration \cite{AkihamaOE2011}. Although entanglement can be generated through protocols that are robust to noise, imperfect local operations and gates add infidelity to the Bell states. To improve the quality of entanglement, it is possible to apply entanglement distillation if two nodes A and B share two Bell states \cite{Bennett1996PRL, Deutsch1996PRL}. A standard entanglement purification protocol \cite{Deutsch1996PRL} can be directly implemented through the parity measurement described above enabling the generation of high-fidelity Bell states (see SM).

Once entanglement has been generated between nodes it is a resource enabling the construction of larger networks. As a particular example, entanglement swapping by measuring the qubits within a node in the Bell basis can be used in a quantum repeater architecture to achieve long distance quantum communications \cite{Briegel1998PRL}. In the proposed circuits, this can be achieved by measuring parity followed by a rotation and a projective measurement of the spins (see SM). Here, the projective spin measurement can be performed by scattering photons off the $\sigma_+$ polarized optical transition from an external laser and using the on-chip detectors to measure the fluorescence.

\begin{figure}[!t]
\includegraphics[width=\columnwidth]{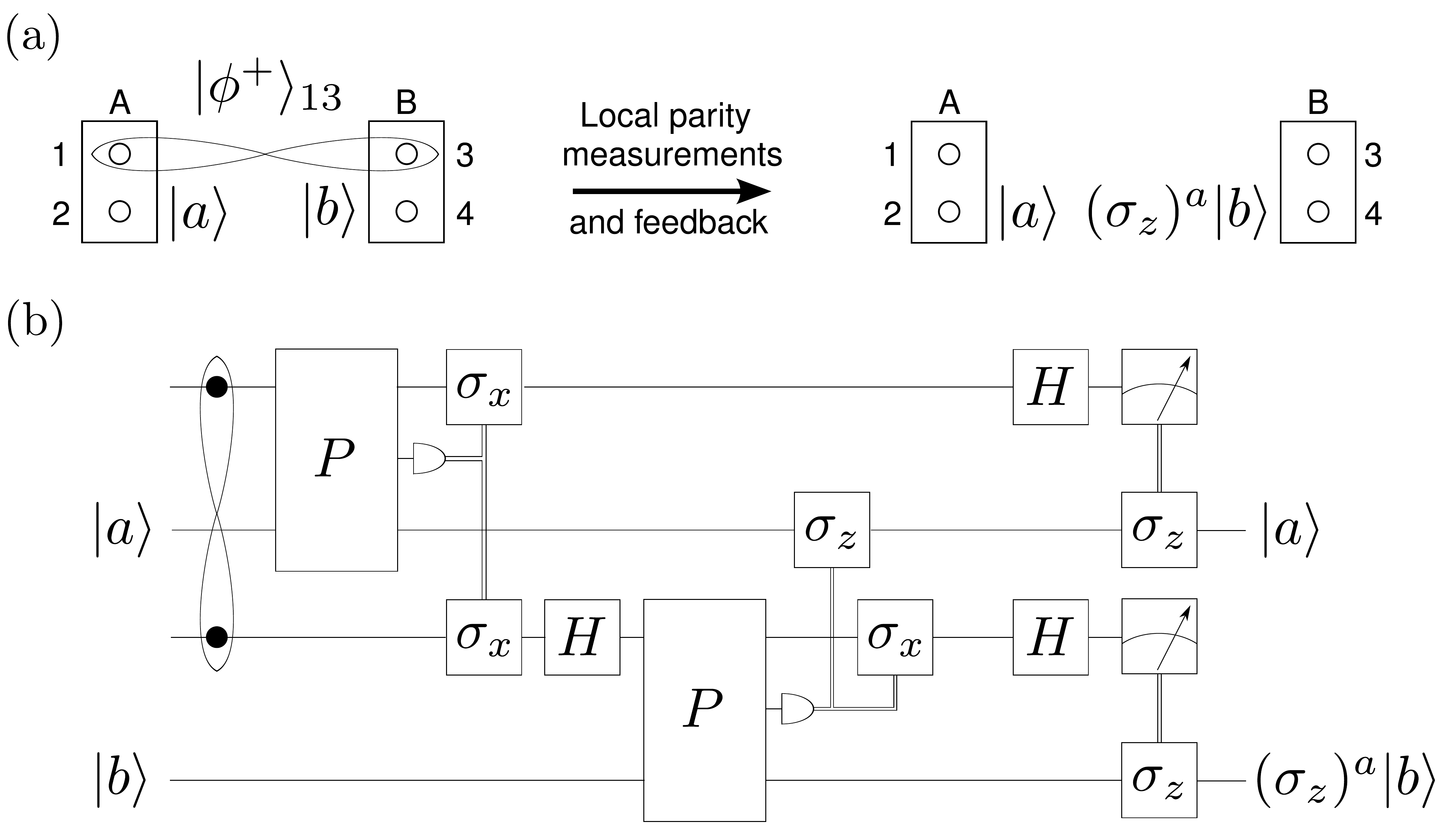}
\caption{\label{fig:CZ} Teleportation-based CZ gate using parity measurements. (a) Schematic showing that a Bell pair $|\phi^+\rangle_{13} = (| 0 0 \rangle + | 1 1 \rangle)/\sqrt{2}$ shared between nodes  can be used as a resource to perform a teleportation-based CZ gate between distant qubits $|a \rangle$ and $| b \rangle$. (b) Circuit diagram for the teleportation-based CZ gate using parity measurements $P$ along with unitary single qubit rotations represented by the $\sigma_x$ and $\sigma_z$ Pauli operators conditioned on the measurement outcomes along with Hadamard transformations $H$.}
\end{figure}

Alternatively, we can use the generated entanglement between nodes to perform a teleportation-based controlled phase (CZ) gate between distant logical qubits $| a \rangle$ and $| b \rangle$, as shown in Fig.~\ref{fig:CZ}(a). In this scheme one emitter in the MZI plays the role of a resource qubit, forming half of a Bell pair distributed across two nodes, while the other is the logical qubit. The circuit diagram of the CZ gate is shown in Fig.~\ref{fig:CZ}(b), and consists of two parts: (i) accumulation of a controlled phase and (ii) a quantum eraser. As shown in the SM, this procedure implements a gate on the logical qubits after successful parity and resource-qubit measurements. The first part involves a parity measurement on each pair at A and B with conditional qubit rotations depending on the measurement outcomes. The quantum eraser then consists of Hadamard rotations $H$ and projective measurements on the resource qubits 1 and 3 with conditional rotations on the logical qubits. Since the parity measurements each operate on different qubits, the fidelity of the CZ gate scales with the product of the two parity measurement fidelities, i.e. ${\cal F}_{\rm CJ}^2$  (see SM). The construction of a gate between distant qubits along these lines is known to be a sufficient resource for universal quantum computation despite it being probabilistic, since it enables the construction of cluster states \cite{Duan2005PRL, Barrett2005PRA}. At the same time it is conceivable that more efficient architectures may be constructed by combining it with error-correcting codes designed to correct for qubit loss \cite{Ralph2005PRL}.

We finally consider implementations based on different quantum emitters and photonic platforms. Quantum dots in PCWs have been used to demonstrate near-unity $\beta$-factors \cite{Arcari2014PRL} with Purcell-enhanced optical transition lifetimes of $\sim \SI{0.2}{\nano\second}$ \cite{Lodahl2015RMP, Arcari2014PRL}. Our implementation is based on coherently scattering photons, therefore, to maintain near-unity fidelity, the single photons need to have temporal widths that are much longer than the emission lifetime, i.e. $\Gamma/\sigma \gtrsim 50$ (see SM).  This means that the optical transition must be free of noise on this time scale. Indistinguishable photon generation through coherent scattering \cite{Matthiesen2013NCOM} indicates that this is possible with QDs. The photon pulse width sets the minimum time required to measure parity. Since the two-click protocol requires two scattered photons, the minimum measurement time is $\gtrsim \SI{20}{\nano\second}$ setting a minimum bound on the required coherence time of the spin states of the qubits. Recent work on hole spins in QDs \cite{Brunner2009Science, Delteil2015arXiv} is encouraging, but further work is required to examine these properties when the external magnetic field points along the growth direction, which is required to suppress diagonal transitions and make the transition dipoles circularly polarized \cite{DreiserPRB2008}. Additionally, an integrated photon source that produces photons of the required linewidth is needed. This may be achieved using a QD whose spontaneous emission lifetime is suppressed \cite{Wang2011PRL} in comparison to the QDs in the MZI, which can be Purcell enhanced. Alternatively, photons may be produced using Raman fluorescence \cite{Fernandez2009PRL}. Finally, our protocol requires coherent spin manipulation which can also be achieved optically or by using microwave pulses matching the detuning of the spin states. Optical spin manipulation can be implemented in QD molecules, which possess both spin-pumping and cycling transitions \cite{Delley2015arxiv}. Alternatively, microwave spin resonances have been demonstrated with QDs \cite{Kroner2008PRL}, but coherent manipulation has yet to be achieved.

In conclusion, we have designed and analysed a simple architecture for a quantum network based on quantum dots. The architecture uses chiral interaction to construct a simple on-chip universal building block which can be merged and scaled to perform universal quantum computation. Our implementation uses stationary qubits for computation and photonic flying qubits for communication and as ancillas to herald gate success.  Although we mainly considered QD-based applications, other emitters also have the potential to be used to construct our proposed circuit. Diamond color centres provide an alternative solid-state platform for realizing chiral photonic systems and can be integrated into photonic nanostructures \cite{Faraon2012PRL}. In particular, silicon vacancy centres' bright zero-phonon line \cite{Pingault2014PRL, Rogers2014PRL} makes them a promising candidate. Efficient atom-based chiral light--matter interaction has also recently been demonstrated \cite{Junge2013PRL} making single atoms trapped near photonic nanostructures another potential platform for realizing our proposed gate.

Sahand Mahmoodian acknowledges discussions with Immo S\"{o}llner, Leonardo Midolo and Alisa Javadi. We gratefully acknowledge financial support from the Villum Foundation, the Carlsberg Foundation, the Danish Council for Independent Research (Natural Sciences and Technology and Production Sciences), and the European Research Council (ERC Consolidator Grants ALLQUANTUM and QIOS).

\bibliography{bigBib}

\appendix
\onecolumngrid
\newpage

\section{Supplementary Information}

\section{The Mach-Zehnder Interferometer for parity measurements}

\begin{figure}[!th]
\includegraphics[width=0.8\columnwidth]{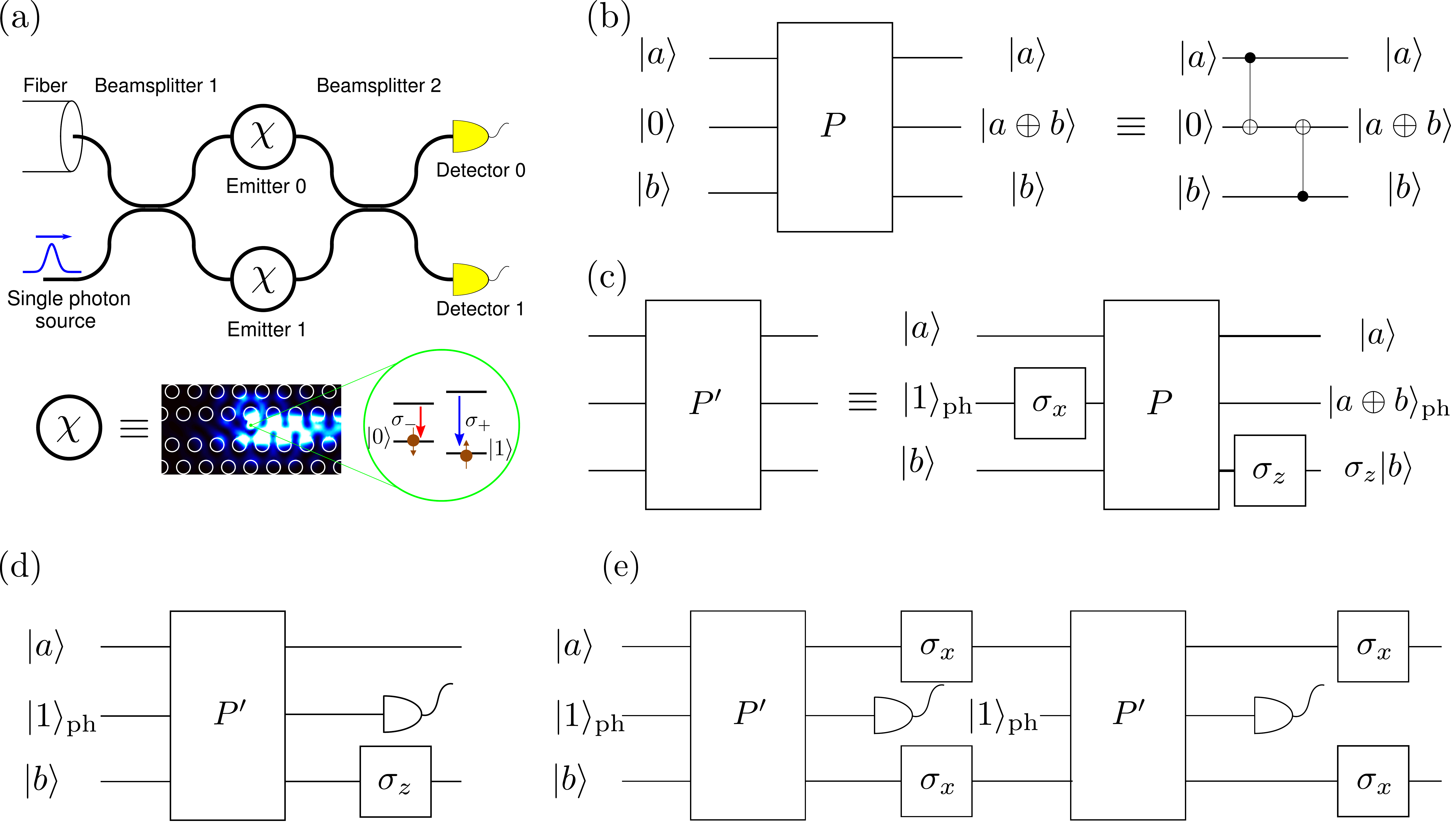}
\caption{\label{fig:MZIparity} (a) Schematic of a photonic circuit for measuring the parity of two qubits positioned in separate arms of a MZI.  (b) Circuit diagram defining the parity measurement in terms of two cNOT operations on the ancilla. (c) Launching and scattering a photon in the MZI in (a) implements a two-qubit parity measurement, which we denote $P'$. It differs from the standard parity measurement $P$ by single qubit rotations. (d) The logical circuit diagram for the one-click protocol. (e) Same as (d), but for the two-click protocol.}
\end{figure}

Here we provide more details concerning the operation of the Mach-Zehnder interferometer (MZI) discussed in the main text. Figure \ref{fig:MZIparity}(a) shows the interferometer. The parity measurement is shown in Fig.~\ref{fig:MZIparity}(b) and can be described by two controlled NOT (cNOT) operations on the ancilla qubit.  Launching and scattering a photon in our MZI implements the parity measurement $P'$, which differs from standard parity measurement $P$ by a single qubit rotation  as shown in Fig.~\ref{fig:MZIparity}(c). The one and two-click protocols use the parity measurement $P'$ and are shown as circuit diagrams in Figs.~\ref{fig:MZIparity}(d)-(e) respectively.   The one-click protocol has a further $\sigma_z$ operation that corrects the phase of the output state, while the two-click protocol has extra $\sigma_x$ operations to improve the measurement fidelity and to restore the qubits to their original state.

We now model the dynamics of the system, firstly for the one-click protocol in the ideal case where the emitters are perfectly coupled to the waveguides and are on resonance with the incident photon, and later we consider losses and detuning for both protocols. In the ideal limit, the emitters' transmission coefficients are $t_0=t_1=-1$. Considering an arbitrary spin-state of the emitters and the photon entering the bottom arm, the initial state is
\begin{equation}
|\textrm{in} \rangle = \left[ c_{00} | 00 \rangle + c_{01} | 01 \rangle + c_{10} | 10 \rangle + c_{11} | 11 \rangle \right] | 1 \rangle_{\rm ph},
\end{equation}
where $|ab\rangle$ means that the top qubit is in state $|a\rangle$ and the bottom is in state $|b\rangle$. For the photonic state we use the notation $|0\rangle_{\rm ph}$ for one photon in the top arm and no photon in the bottom arm, and $|1\rangle_{\rm ph}$ for no photon in the top arm and one photon in the bottom arm. The beamsplitter couples the photonic modes through the transformation
\begin{equation}
\begin{split}
|0 \rangle_{\rm ph} &\rightarrow \frac{1}{\sqrt{2}} \left( |0 \rangle_{\rm ph} + i | 1 \rangle_{\rm ph} \right)\\
|1 \rangle_{\rm ph} &\rightarrow \frac{1}{\sqrt{2}} \left( i|0 \rangle_{\rm ph} +  | 1 \rangle_{\rm ph} \right),
\end{split}
\end{equation}
and therefore after the first beamsplitter the system is in the state
\begin{equation}
\frac{1}{\sqrt{2}} \left[ c_{00} | 00 \rangle + c_{01} | 01 \rangle + c_{10} | 10 \rangle + c_{11} | 11 \rangle \right]  \left( i|0 \rangle_{\rm ph} +  | 1 \rangle_{\rm ph} \right).
\end{equation}
The photon scatters off the emitter only if it is in $|1\rangle$, i.e. spin up. After the photon scatters off the emitters and exits through Beamsplitter 2, the output state is
\begin{equation}
i |0 \rangle_{\rm ph} \left( c_{00} | 00 \rangle - c_{11} | 1 1 \rangle \right)  + |1 \rangle_{\rm ph} \left( -c_{01} | 01 \rangle + c_{10} | 1 0 \rangle \right),
\end{equation}
which is followed by a $\sigma_z$ operation on the second qubit returning the desired output
\begin{equation}
|\textrm{out} \rangle = i |0 \rangle_{\rm ph} \left( c_{00} | 00 \rangle + c_{11} | 1 1 \rangle \right)  + |1 \rangle_{\rm ph} \left(c_{01} | 01 \rangle + c_{10} | 1 0 \rangle \right).
\end{equation}
The photonic circuit and $\sigma_z$ operation thus implement a quantum non-demolition parity measurement on the two spin states.

\section{One and two-click protocols}
\label{sec:12prot}

We now consider the operation of the above gate in the presence of an imbalanced interferometer due to emitters having different directional beta-factors $\beta$ and different detunings $\Delta$. The transmission coefficient for single photon scattering as a function of $\beta$ and $\Delta$ is
\begin{equation}
\label{eq:transCoeff}
t = 1-\frac{2 \beta}{1 - 2i \beta \Delta /\Gamma_R},
\end{equation}
where $\Gamma_R$ is the coupling rate to the right-propagating waveguide mode. When $\beta<1$ the transmission coefficient is less than unity $|t|<1$, and when $\Delta \neq 0$ the phase of the transmission coefficient differs from the ideal value $\arg(t)\neq\pi$. Both effects act to unbalance the interferometer and spoil the gate operation. Here we consider the effect of this on both the one and two-click protocols.

Depending on the magnitude of loss and detuning, the parity measurement may be performed using the one-click or two-click protocol. The one-click protocol requires launching and detecting a single photon, while for the two-click protocol,  $\sigma_x$ operations are performed on the two stationary qubits after the first detection and a second photon is launched and detected. As we now show, if the loss and detuning are small, the infidelity for the one-click protocol scales with $\epsilon^2$, while for the two click protocol it scales with $\epsilon^4$, where $\epsilon$ is the fraction of loss or detuning.

We start with an arbitrary pure input state
\begin{equation}
| \textrm{in} \rangle = \left[ c_{00} | 0 0 \rangle + c_{01} |0 1 \rangle + c_{10} | 1 0 \rangle + c_{11} | 1 1 \rangle \right] |1 \rangle_{\rm ph}.
\end{equation}
After the second beam splitter the final state before photon detection is
\begin{equation}
\begin{split}
| \textrm{out} \rangle^{(1)} = | 0 \rangle_{\rm ph} &\left[ c_{00} | 0 0 \rangle +\frac{1+t_1}{2}c_{01} |0 1 \rangle + \frac{1+t_0}{2} c_{10} | 1 0 \rangle +\frac{t_0+t_1}{2} c_{11} |11\rangle \right]\\
+ & | 1 \rangle_{\rm ph} \left[ \frac{t_1-1}{2}c_{01} |0 1 \rangle + \frac{1-t_0}{2} c_{10} | 1 0 \rangle + \frac{t_0-t_1}{2} c_{11} |11\rangle \right] + |\textrm{lost} \rangle,
\end{split}
\end{equation}
where $t_0$ and $t_1$ are the scattering coefficients of the top and bottom emitters respectively. As expected $| 0 0 \rangle$ and $| 1 1 \rangle$ lead to the photon state $| 0 \rangle_{\rm ph}$, and $| 0 1 \rangle$ and $| 1 0 \rangle$ lead to the photon state $| 1 \rangle_{\rm ph}$ for ideal scattering $t_0=t_1=-1$, but $|01 \rangle$, $| 1 0 \rangle$, and $| 1 1 \rangle$ can also lead to erroneous detection events due to imperfect scattering. These errors contribute to lowering the fidelity of the output state. A click on Detector 0 followed by a $\sigma_z$ operation on the second qubit leaves the state as
\begin{equation}
\begin{split}
| \textrm{out} \rangle^{(1)}_{\rm click\, 0} =  \left[ c_{00} |0 0 \rangle - \frac{1+t_1}{2} c_{01} | 0 1 \rangle + \frac{1+t_0}{2} c_{10} | 1 0 \rangle - \frac{t_0+t_1}{2} c_{11} | 1 1 \rangle \right]/\sqrt{P_{\rm click\,0}^{(1)}},
\end{split}
\end{equation}
where the probability to get this outcome is given by $P_{\rm click\,0}^{(1)} = |c_{00}|^2 + |1+t_1|^2|c_{01}|^2/4 + |1+t_0|^2 |c_{10}|^2/4 +|t_0+t_1|^2 |c_{11}|^2/4$. The fidelity of the output state is defined as ${\cal F}_{\rm click \, 0} = | \langle \textrm{ideal} | \textrm{out} \rangle^{(1)}_{\rm click\, 0}|^2$, where $| \textrm{ideal} \rangle = (c_{00} | 0 0 \rangle + c_{11} | 1 1 \rangle) / \sqrt{|c_{00}|^2 + |c_{11}|^2}$ Alternatively, a click on Detector 1 followed by a $\sigma_z$ operation leaves the state as
\begin{equation}
\begin{split}
| \textrm{out} \rangle^{(1)}_{\rm click\, 1} = \left[ \frac{1-t_1}{2}c_{01} |0 1 \rangle + \frac{1-t_0}{2} c_{10} | 1 0 \rangle - \frac{t_0-t_1}{2} c_{11} |11\rangle \right] /\sqrt{P_{\rm click\,1}^{(1)}},
\end{split}
\end{equation}
where the probability of this event is $P_{\rm click\,1}^{(1)} = |1-t_1|^2|c_{01}|^2/4 + |1-t_0|^2 |c_{10}|^2/4 +|t_0-t_1|^2 |c_{11}|^2/4$. Here the ideal output state is $| \textrm{ideal} \rangle = (c_{01} | 0 1 \rangle + c_{10} | 1 0 \rangle) / \sqrt{|c_{01}|^2 + |c_{10}|^2}$ with the conditional fidelity ${\cal F}_{\rm click \, 1} = | \langle \textrm{ideal} | \textrm{out} \rangle^{(1)}_{\rm click\, 1}|^2$. The fidelity of the parity measurement conditioned on a single click on either Detector 0 or 1 is defined as
\begin{equation}
\label{eq:fidelity1click}
{\cal F}^{\rm \, 1 \, click} = \frac{P_{\rm click\,0}^{(1)} {\cal F}_{\rm click \, 0} + P_{\rm click\,1}^{(1)} {\cal F}_{\rm click \, 1}}{P_{\rm click\,0}^{(1)} + P_{\rm click\,1}^{(1)}}.
\end{equation}

The success of the two-click protocol is conditioned on two clicks on Detector 0 or Detector 1. We first consider two clicks on Detector 0. After the first click on Detector 0, instead of performing a $\sigma_z$ operation, we perform a $\sigma_x$ operation on both qubits leaving the state as
\begin{equation}
\begin{split}
| \textrm{out} \rangle^{(1)}_{\rm click\, 0} = \left[ c_{00} |1 1 \rangle + \frac{1+t_1}{2} c_{01} | 1 0 \rangle + \frac{1+t_0}{2} c_{10} | 0 1 \rangle + \frac{t_0+t_1}{2} c_{11} | 0 0 \rangle \right]/\sqrt{P_{\rm click\,0}^{(1)}},
\end{split}
\end{equation}
After scattering the second photon, the output state is
\begin{equation}
\begin{split}
|\textrm{out} \rangle^{(2)}_{\rm click\, 0} = &|0\rangle_{\rm ph} \left[ \frac{t_0+t_1}{2}\left(c_{00} |11\rangle + c_{11} |00\rangle \right) + \frac{(1+t_0)(1+t_1)}{4}\left(c_{10} |01\rangle + c_{01} |10\rangle \right) \right]/\sqrt{P_{\rm click\,0}^{(1)}}\\
& + | 1 \rangle{\rm ph} \left[ \frac{(t_1-1)(1+t_0)}{4} c_{10} | 0 1 \rangle + \frac{(1-t_0)(1+t_1)}{4} c_{10} | 0 1 \rangle + \frac{t_1-t_0}{2}c_{00} |11\rangle \right]/\sqrt{P_{\rm click\,0}^{(1)}} + | \textrm{lost} \rangle,
\end{split}
\end{equation}
which, after a second click on Detector 0 followed by $\sigma_x$ operations to return the qubits to their original state, becomes
\begin{equation}
\begin{split}
\label{eq:out00}
|\textrm{out} \rangle^{(2)}_{\rm click \, 0,0} = \left[ \frac{t_0+t_1}{2}\left(c_{00} |00\rangle + c_{11} |11\rangle \right) + \frac{(1+t_0)(1+t_1)}{4}\left(c_{10} |10\rangle + c_{01} |01\rangle \right) \right]/\sqrt{P_{\rm click\,0,0}^{(2)}}.
\end{split}
\end{equation}
The probability of the error terms now scales quartically since there are now two terms  $(1+t_0)$ and $(1+t_1)$, which vanish in the ideal limit $t_0=t_1=-1$. This comes at the expense of a reduced success probability. The probability of getting two successive clicks on Detector 0 is
\begin{equation}
\label{eq:Pclick00}
P_{\rm click\,0,0}^{(2)}  = \frac{1}{4}|t_0 + t_1|^2 (|c_{00}|^2 + |c_{11}|^2) + \frac{1}{16}|1 + t_1|^2 |1 + t_0|^2 (|c_{01}|^2 + |c_{10}|^2).
\end{equation}
The fidelity of this outcome is
\begin{equation}
\label{eq:Fclick00}
{\cal F}_{\rm click \, 0,0} = \frac{|t_0+t_1|^2 (|c_{00}|^2+|c_{11}|^2)}{4 P_{\rm click\,0,0}^{(2)}}.
\end{equation}
Carrying out the two-click protocol for two clicks on Detector 1 followed by $\sigma_x$ operations leads to a final output state
\begin{equation}
\label{eq:out11}
|\textrm{out} \rangle^{(2)}_{\rm click \, 1,1} = \frac 1 4 (1-t_0)(t_1-1) (c_{10}|10\rangle + c_{01} | 01 \rangle)/ \sqrt{P_{\rm click\,1,1}^{(2)}},
\end{equation}
with probability $P_{\rm click\,1,1}^{(2)} = |1-t_0|^2 |1-t_1|^2 (|c_{01}|^2 + |c_{10}|^2)/16$. Here, the fidelity is unity ${\cal F}_{\rm click \, 1,1}=1$ and is thus insensitive to loss, detuning, and differences in the transmission coefficients. We define the average fidelity of the parity measurement for the two-click protocol as
\begin{equation}
\label{eq:fidelity}
{\cal F}^{\rm 2 \, click} = \frac{P_{\rm click\,0,0}^{(2)} {\cal F}_{\rm click \, 0,0} + P_{\rm click\,1,1}^{(2)} {\cal F}_{\rm click \, 1,1}}{P_{\rm click\,0,0}^{(2)} + P_{\rm click\,1,1}^{(2)}}.
\end{equation}

\subsection{Choi-Jamiolkowski Fidelity}

We now use the above expressions for the input-state-dependent fidelities ${\cal F}^\textrm{\,1 click}$ and ${\cal F}^\textrm{\,2 click}$ to determine the fidelity of our parity operation. If an ideal parity operation is represented by the superoperator ${\cal O}$, and our potentially imperfect implementation is denoted ${\cal A}$, it is important to know how close these two operations are. It has been shown that a characterization of an operation can be obtained through the Choi-Jamiolkowski fidelity ${\cal F}_{\rm CJ}$ \cite{Gilchrist2005PRA}. In general, to compute this, we consider the $d$-dimensional maximally entangled state $| \Phi \rangle = \sum_j |j \rangle | j \rangle/ \sqrt{d}$ in an orthonormal basis $\{| j \rangle\}$ as an input state. We then compute
\begin{equation}
\begin{split}
&\rho_{\cal O} \equiv \left[ {\cal O} \otimes {\cal I} \right] (| \Phi \rangle \langle \Phi |),\\
&\rho_{\cal A} \equiv \left[ {\cal A} \otimes {\cal I} \right] (| \Phi \rangle \langle \Phi |),
\end{split}
\end{equation}
i.e. the operators operate on one part of the entangled state while the other is acted on by the identity superoperator ${\cal I}$. The fidelity of these two states gives ${\cal F}_{\rm CJ}$, which is straightforward to compute as $\rho_{\cal O}$ is a pure state after a successful photon detection.

For two-qubit gate operations, computing the CJ fidelity is closely related to entanglement swapping: consider two Bell pairs $| \phi^+ \rangle_{12} | \phi^+ \rangle_{34}$, where we define $| \phi^\pm \rangle = (| 0 0 \rangle \pm | 1 1 \rangle)/\sqrt{2}$ and $| \psi^\pm \rangle = (| 0 1 \rangle \pm | 1 0 \rangle)/\sqrt{2}$. Entanglement can be swapped by measuring qubits 2 and 3 in the Bell basis, and here we also measure the parity of qubits 2 and 3 and then compute the fidelity of the resulting state with the ideal two-qubit operation. Rewriting the CJ state as
\begin{equation}
|\Phi\rangle = | \phi^+ \rangle_{12} | \phi^+ \rangle_{34} = \frac 1 2 \left[ |\phi^+ \rangle_{23} |\phi^+\rangle_{14} + |\phi^- \rangle_{23} |\phi^-\rangle_{14} + |\psi^+ \rangle_{23} |\psi^+\rangle_{14} + |\psi^- \rangle_{23} |\psi^-\rangle_{14} \right],
\end{equation}
we can compute the CJ fidelity for the parity measurement using the above input state and the definitions in (\ref{eq:fidelity1click}) and (\ref{eq:fidelity}). The one-click protocol has fidelity
\begin{equation}
\label{eq:CJfidelity1click}
{\cal F}^{\rm \, 1 \, click}_{\rm CJ} = \frac{|1-\frac{t_0+t_1}{2}|^2}{1+(|t_0+t_1|^2+|1+t_1|^2+|1+t_0|^2+|1-t_0|^2+|1-t_1|^2+|t_0-t_1|^2)/4},
\end{equation}
while the two-click protocol has fidelity
\begin{equation}
{\cal F}_{\rm CJ}^{\rm 2 \, click} = 1- \frac{|1+t_1|^2|1+t_0|^2}{|1+t_1|^2|1+t_0|^2 + |1-t_1|^2|1-t_0|^2 + 4 |t_0+t_1|^2}.
\end{equation}
When $\Delta_0=\Delta_1=0$ and $\beta_0=\beta_1=\beta$, ${\cal F}^{\rm \, 1 \, click}_{\rm CJ} = 1- (1- \beta)^2/(1 - 2 \beta + 2 \beta^2)$ and ${\cal F}^{\rm \, 1 \, click}_{\rm CJ} = 1 - (1-\beta)^4/[\beta^4 + (1-2\beta)^2 + (1-\beta)^4]$.

We also use the Choi-Jamiolkowski input state to compute the typical success probability for the two protocols. When $\Delta=0$ and $\beta_0=\beta_1=\beta$, the success probabilities are
\begin{equation}
P_{\rm success}^{\rm \, 1 \, click} = \frac 1 4 \left[1+ (1-2\beta)^2 + 2 \beta^2 + 2(1-\beta)^2 \right],
\end{equation}
and
\begin{equation}
P_{\rm success}^{\rm \, 2 \, click} = \frac 1 2 \left[(1-2 \beta)^2 + \beta^4 + (1-\beta)^4 \right].
\end{equation}

\subsection{Finite-width photon pulses}

We have thus far considered photon pulses which are infinitely narrow in frequency. Here we relax this assumption and compute the resulting infidelity for the two-click protocol due to a finite photon bandwidth. We denote the input state as
\begin{equation}
|\textrm{in}\rangle =  \left( c_{00} |0 0 \rangle + c_{01} | 0 1 \rangle + c_{10} | 1 0 \rangle + c_{11}| 1 1 \rangle  \right) \int dk \, f(k) \hat{a}_{k, 1}^{\dagger} | \varnothing \rangle,
\end{equation}
where $\omega = c \, k$ is the photon angular frequency and $c$ is the speed of light in vacuum, $| \varnothing \rangle$ is the vacuum state, and $f(k)$ is the functional form of the photon wavepacket. The operator $\hat{a}_{k, i}^{\dagger}$ creates a photon with frequency $k c$ in mode $i=0,1$. Here we consider Lorentzian pulses of the form
\begin{equation}
\label{eq:lorentzian}
f(k) = \frac{\sqrt{2 \sigma^3 / \pi}}{\sigma^2 + c^2k^2}.
\end{equation}
Using the scattering coefficient from (\ref{eq:transCoeff}), the output state after scattering the first photon is
\begin{equation}
\begin{split}
| \textrm{out} \rangle^{(1)} = \int dk \,  f(k) \,  & | 0_k \rangle_{\rm ph} \left(  c_{00} | 00 \rangle + \frac 1 2 (t_k^1 + 1)c_{01} | 0 1 \rangle + \frac 1 2 (t_k^0 + 1)c_{10} | 1 0 \rangle + \frac 1 2 (t_k^0 + t_k^1) c_{11} | 11 \rangle \right)\\
 +\int dk \, f(k) \, & | 1_k \rangle_{\rm ph} \left( \frac 1 2 (t_k^1 - 1) c_{01} | 0 1 \rangle + \frac 1 2 (1- t_k^0) c_{10} | 1 0 \rangle  + \frac 1 2 (t_k^0 - t_k^1) c_{11} | 1 1 \rangle \right) + | \textrm{lost} \rangle,
\end{split}
\end{equation}
where $| i_k \rangle_{\rm ph}$ is a single photon in mode $i$ with wavevector $k$. A click on Detector 0 followed by the $\sigma_z$ operation projects the state into
\begin{equation}
\begin{split}
\label{eq:out0finitesigma}
| \textrm{out} \rangle^{(1)}_\textrm{click 0} = \int dk \,  f(k) \,  & | 0_k \rangle_{\rm ph} \left(  c_{00} | 00 \rangle - \frac 1 2 (t_k^1 + 1)c_{01} | 0 1 \rangle + \frac 1 2 (t_k^0 + 1)c_{10} | 1 0 \rangle - \frac 1 2 (t_k^0 + t_k^1) c_{11} | 11 \rangle \right)/\sqrt{P_\textrm{click 0}^{(1)}},
\end{split}
\end{equation}
where
\begin{equation}
P_\textrm{click 0}^{(1)} = \int dk \, |f(k)|^2 \left[ |c_{00}|^2 + \frac 1 4 |t_{k}^0 + t_{k}^1|^2 |c_{11}|^2 + \frac 1 4 |t_k^1+1|^2 |c_{01}|^2 + \frac 1 4 |t_k^0+1|^2 |c_{10}|^2 \right].
\end{equation}
Alternatively a click on Detector 1 and the subsequent $\sigma_z$ operation leaves the state as
\begin{equation}
\begin{split}
\label{eq:out1finitesigma}
| \textrm{out} \rangle^{(1)}_\textrm{click 1} = \int dk \, f(k) \, & | 1_k \rangle_{\rm ph} \left( \frac 1 2 (1 - t_k^1) c_{01} | 0 1 \rangle + \frac 1 2 (1- t_k^0) c_{10} | 1 0 \rangle  - \frac 1 2 (t_k^0 - t_k^1) c_{11} | 1 1 \rangle \right)/\sqrt{P_\textrm{click 1}^{(1)}},
\end{split}
\end{equation}
where
\begin{equation}
P_\textrm{click 1}^{(1)} = \int dk \, |f(k)|^2 \left[ \frac 1 4 |t_{k}^1 - 1|^2 |c_{01}|^2 + \frac 1 4 |1 - t_k^0|^2 |c_{10}|^2 + \frac 1 4 |t_k^0-1|^2 |c_{11}|^2 \right].
\end{equation}
Using (\ref{eq:fidelity1click}), (\ref{eq:out0finitesigma}), (\ref{eq:out1finitesigma}), and tracing over the photon states, we can compute the CJ fidelity for the one-click protocol for photon wave packets. In the limit $\beta_0 \rightarrow 1$, $\beta_1 \rightarrow 1$, the CJ fidelity is
\begin{equation}
\begin{split}
\mathcal{F}_{\rm CJ}^\textrm{1 click} = 1- \sigma^2 \left( \frac{\Gamma_0 + \Gamma_1}{\Gamma_0 \Gamma_1} \right)^2.
\end{split}
\end{equation}

For the two-click protocol, the output state after two clicks on Detector 0 with $\sigma_x$ operations after the first and second clicks is
\begin{equation}
\begin{split}
\label{eq:out00finitesigma}
| \textrm{out} \rangle^{(2)}_\textrm{click 0,0} = &\int dk \, dk' f (k) f(k') | 0_k \rangle_{\rm ph} | 0_{k'} \rangle_{\rm ph} \left[ \frac 1 4 (1+t_k^1) (1+t_{k'}^0) c_{01} | 0 1 \rangle + \frac 1 4 (1+t_k^0) (1+t_{k'}^1) c_{10} | 1 0 \rangle \right. \\
&+ \left. \frac 1 2 (t_{k'}^0 + t_{k'}^1) c_{00} | 0 0 \rangle + \frac 1 2 (t_k^0 + t_k^1) c_{11} | 1 1 \rangle \right] /\sqrt{P_\textrm{click 0,0}^{(2)}},
\end{split}
\end{equation}
where
\begin{equation}
\begin{split}
P_{\rm click 0,0}^{(2)} = &\int dk \, dk' \, |f(k)|^2 |f(k')|^2 \left[ \frac 1 4 |t_{k'}^0 + t_{k'}^1|^2 |c_{00}|^2 + \frac 1 4 |t_{k}^0 + t_{k}^1|^2 |c_{11}|^2 + \frac{1}{16} |1+t_k^1|^2 |1+t_{k'}^0|^2 |c_{01}|^2 \right. \\
& \left. + \frac{1}{16} |1+t_{k'}^1|^2 |1+t_{k}^0|^2 |c_{10}|^2 \right].
\end{split}
\end{equation}
The output state after two clicks on Detector 1 is
\begin{equation}
\begin{split}
\label{eq:out11finitesigma}
| \textrm{out} \rangle^{(2)}_\textrm{click 1,1} = &\int dk \, dk' f (k) f(k') | 1_k \rangle_{\rm ph} | 1_{k'} \rangle_{\rm ph} \left[ \frac 1 4 (1-t_{k'}^0) (t_{k}^1-1) c_{01} | 0 1 \rangle + \frac 1 4 (1-t_k^0) (t_{k'}^1-1) c_{10} | 1 0 \rangle \right] /\sqrt{P_{\rm click 1,1}^{(2)}},
\end{split}
\end{equation}
where
\begin{equation}
\begin{split}
P_{\rm click 1,1}^{(2)} = \frac{1}{16}\int dk \, dk' |f (k)|^2 |f(k')|^2 \left[ |t_k^1-1|^2|1-t_{k'}^0|^2 |c_{01}|^2 + |t_{k'}^1-1|^2|1-t_{k}^0|^2 |c_{10}|^2 \right].
\end{split}
\end{equation}
As for the one-click protocol we compute the CJ fidelity from (\ref{eq:fidelity}), (\ref{eq:out00finitesigma}), (\ref{eq:out11finitesigma}), and trace over the photon states. In the limit $\beta_0 \rightarrow 1$, $\beta_1 \rightarrow 1$, the CJ fidelity scales as
\begin{equation}
\begin{split}
\mathcal{F}_{\rm CJ}^\textrm{2 click} = 1- 2 \sigma^2 \left( \frac{1}{\Gamma_0^2} + \frac{1}{\Gamma_1^2} \right).
\end{split}
\end{equation}
Unlike for single frequency photons the two-click protocol does not improve the scaling of the fidelity over the one-click protocol. This is because the finite width of the photons causes entanglement between the photon state and the qubit states leaving ``which-way'' information which degrades the quantum state. This is clear from examining the coefficients of the qubits in (\ref{eq:out00finitesigma}) and (\ref{eq:out11finitesigma}), where the functions of $k'$ and $k$ are different.  Nevertheless, by using sufficiently narrow photon pulses this effect can be overcome. For example for $\Gamma_0^R = \Gamma_1^R = \Gamma_R$ and $\sigma=\Gamma_R/100$, we get $1-\mathcal{F}_{\rm CJ}^\textrm{2 click} = 4\times10^{-4}$, which becomes negligible compared to other effects. This does however set a limit on the time taken to measure the qubits' parity with sufficient accuracy

\subsection{Incoherent scattering}

Thus far, we have considered a directional $\beta$-factor $\beta = \Gamma_R/ (\Gamma_R + \Gamma_L + \gamma_{\rm rad})$, but we now also consider the influence of an incoherent scattering rate $\gamma_{\rm inc}$ due to Markovian dephasing processes. For simplicity, we compute the gate fidelity when $\Delta=0$ and $\beta=1$ ($\Gamma_L = \gamma_{\rm rad}=0$), but $\beta_c<1$, and assume that the two emitters have the same transmission coefficient for coherent scattering $t=-1 + 2 \gamma_{\rm inc}/(\Gamma_{\rm R} + \gamma_{\rm inc})$. We treat incoherent scattering as a phase randomizing process, but ignore the corresponding spectral signature as the photon detectors are typically not sensitive to frequency fluctuations on the order of the emitter linewidth, which simplifies the calculations. The phase randomizing process leads to the photon exiting both arms of the interferometer with equal probability therefore introducing erroneous detection events.

As before, we start with the input state
\begin{equation}
|\textrm{in} \rangle = \left(c_{00} |00\rangle + c_{01} |01 \rangle +c_{10} |10\rangle +c_{11} |11\rangle \right) |1\rangle_{\rm ph}.
\end{equation}
The density matrix of the output state is
\begin{equation}
\begin{split}
\rho_{\rm out 1} = |\textrm{out} \rangle \langle \textrm{out} | + \frac{(1-t^2)}{4} & \left[ (c_{10}| 1 0 \rangle + c_{11} | 1 1 \rangle ) (c_{10}^* \langle 1 0 | +  c_{11} ^* \langle  1 1 | ) + (c_{01}| 0 1 \rangle +  c_{11} | 1 1 \rangle )(c_{01}^* \langle 0 1 | +  c_{11} ^* \langle  1 1 | )\right]\\
& \times \left[ | 0 \rangle_\textrm{ph} \langle 0 | + | 1 \rangle_\textrm{ph} \langle 1 | \right]
\end{split}
\end{equation}
where we have used the notation $|a b \rangle^\dagger= \langle a b |$, and  the output state is split into a pure part
\begin{equation}
|{\rm out} \rangle = | 0 \rangle_{\rm ph} (c_{00} | 00\rangle +\frac 1 2 (1+t) c_{01} | 0 1 \rangle + \frac 1 2 (1+t) c_{10} |1 0 \rangle + t c_{11} | 1 1 \rangle) + | 1 \rangle_{\rm ph} \, \frac 1 2 (t-1) \, (c_{01} | 0 1 \rangle - c_{10} | 1 0 \rangle),
\end{equation}
and a mixed part. As before, we compute the states conditioned on registering a click on Detector 0 or 1, and then compute the fidelity. In the limit where the incoherent fraction is small, i.e. $t \rightarrow -1$, the CJ infidelity is
\begin{equation}
1- {\cal F}^\textrm{ 1 click}_{\rm CJ} \sim \frac{\gamma_{\rm inc}}{\Gamma_{\rm R} + \gamma_{\rm inc}},
\end{equation}
which scales linearly with the fraction of incoherent scattering.

We now compute the output states and fidelity for the two-click protocol.  After registering two clicks on Detector 0 with $\sigma_x$ operations before and after the photon scattering, the density matrix is
\begin{equation}
\begin{split}
\rho_\textrm{out 2}^\textrm{click 0,0} &= \frac {1}{P_\textrm{click 0,0}} |\textrm{out}_2\rangle^\textrm{click 0,0} \langle \textrm{out}_2 | + \frac{(1-t^2)}{4 P_\textrm{click 0,0}}\left(\frac 1 2 (t+1) c_{10} | 1 0 \rangle + c_{11} | 1 1 \rangle \right) \left(\frac 1 2 (t+1) c_{10}^* \langle 1 0 | + c_{11}^* \langle 1 1 | \right)\\
&+ \frac{(1-t^2)}{4 P_\textrm{click 0,0}}\left(\frac 1 2 (t+1) c_{01} | 0 1 \rangle + c_{11} | 1 1 \rangle \right) \left(\frac 1 2 (t+1) c_{01}^* \langle 0 1 | + c_{11}^* \langle 1 1 | \right)\\
&+ \frac{(1-t^2)}{4 P_\textrm{click 0,0}}\left(\frac 1 2 (t+1) c_{10} | 1 0 \rangle + c_{00} | 0 0 \rangle \right) \left(\frac 1 2 (t+1) c_{10}^* \langle 1 0 | + c_{00}^* \langle 0 0 | \right)\\
&+ \frac{(1-t^2)}{4 P_\textrm{click 0,0}}\left(\frac 1 2 (t+1) c_{01} | 0 1 \rangle + c_{00} | 0 0 \rangle \right) \left(\frac 1 2 (t+1) c_{10}^* \langle 0 1 | + c_{00}^* \langle 0 0 | \right)\\
&+ \frac{(1-t^2)}{16 P_\textrm{click 0,0}}\left( |c_{10}|^2 | 1 0 \rangle \langle 0 1| + |c_{01}|^2 | 0 1 \rangle \langle 0 1 | \right),
\end{split}
\end{equation}
where
\begin{equation}
|\textrm{out}_2\rangle^\textrm{click 0,0} = t (c_{00}| 0 0 \rangle + c_{11} | 1 1 \rangle) + \frac 1 4 (t+1)^2 (c_{01} | 0 1 \rangle + c_{10} | 1 0 \rangle ),
\end{equation}
and
\begin{equation}
P_\textrm{click 0,0} = (|c_{00}|^2 + |c_{11}|^2) \left[ t^2 + \frac{1-t^2}{2} \right] + (|c_{01}|^2 + |c_{10}|^2) \left[ \frac{(1+t)^4}{8} + \frac{(1-t^2) (1+t)^2}{16} + \frac{(1-t^2)^2}{16} \right].
\end{equation}
Here, the density matrix is composed of terms corresponding to two coherent photon scattering processing, one coherent and one incoherent scattering process, and two incoherent scattering processes. The output state can also be computed for two clicks on Detector 1,
\begin{equation}
\begin{split}
\rho_\textrm{out 2}^\textrm{click 1,1} &= \frac {1}{P_\textrm{click 1,1}} |\textrm{out}_2\rangle^\textrm{click 1,1} \langle \textrm{out}_2 | + \frac{(1-t^2)(t-1)^2}{8 P_\textrm{click 1,1}} \left( |c_{10}|^2 | 1 0 \rangle \langle 1 0 \rangle + |c_{01}|^2 | 0 1 \rangle \langle 0 1 | \right)\\
& + \frac{(1-t^2)^2}{16 P_\textrm{click 1,1}} \left(  |c_{01}|^2 |  0 1 \rangle \langle 0 1 | + |c_{10}|^2 | 1 0 \rangle \langle 1 0 | \right),
\end{split}
\end{equation}
where
\begin{equation}
|\textrm{out}_2\rangle^\textrm{click 1,1} = \frac 1 4 (t-1)(1-t) (c_{01}| 0 1 \rangle + c_{10} | 1 0 \rangle),
\end{equation}
and
\begin{equation}
P_\textrm{click 1,1} = (|c_{01}|^2 + |c_{10}|^2) \left[ \frac{(1-t)^4}{16} + \frac{(1-t^2)(t-1)^2}{8} + \frac{(1-t^2)^2}{16} \right].
\end{equation}
Using the Choi-Jamiolkowski input state, from these we can compute the CJ fidelity. Again, in the limit $t \rightarrow -1$, the fidelity is
\begin{equation}
1 - {\cal F}_\textrm{CJ}^\textrm{1 click} \sim \frac 5 8 \frac{\gamma_{\rm inc}}{\Gamma_R + \gamma_{\rm inc}},
\end{equation}
which again scales linearly. Therefore, irrespective of whether the one or two-click protocol is used, the infidelity scales linearly with the fraction of incoherent scattering $(1-\beta_c)$.

\section{Entanglement purification}

\begin{figure}[!th]
\includegraphics[width=0.8\columnwidth]{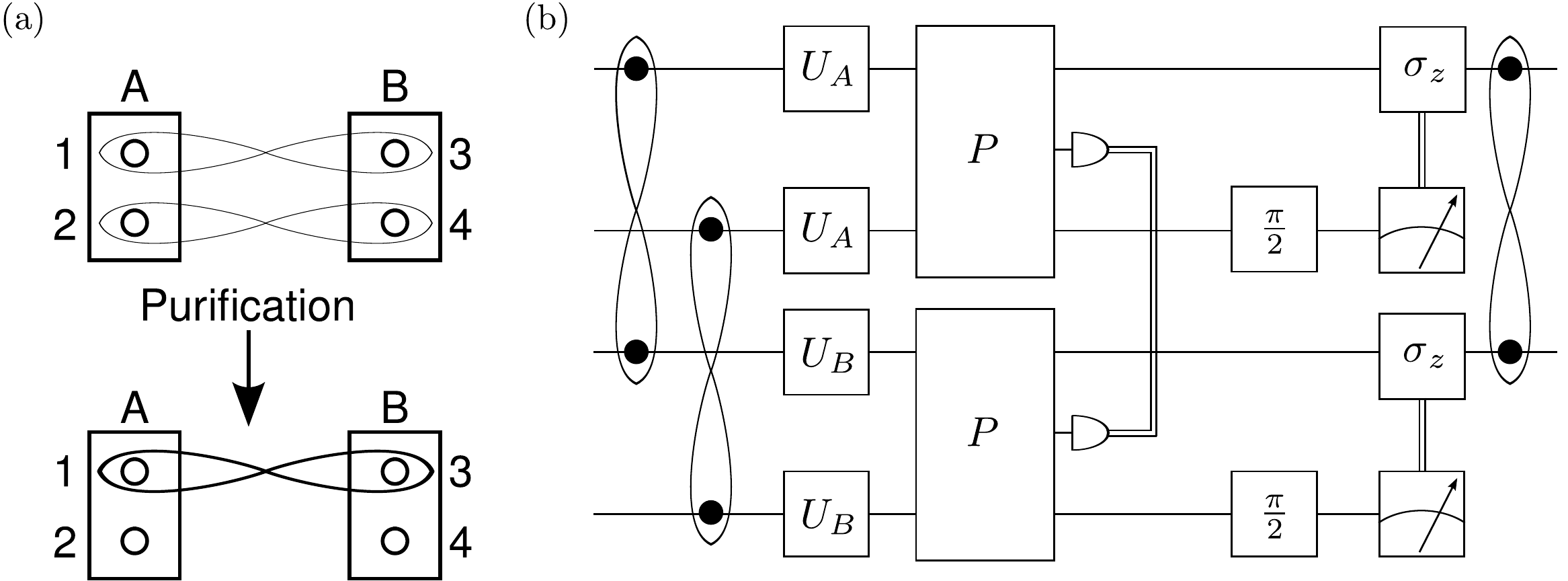}
\caption{\label{fig:purification} Entanglement purification. (a) Schematic showing two entangled pairs of low fidelity being purified to one pair of higher fidelity. (b) Circuit diagram for entanglement purification using parity measurements to distil two impure Bell pairs into one pair of higher fidelity. The operations $U_A$, $U_B$, and $\pi/2$ are unitary operations (see text). The double lines joining the photodetectors indicate communication between A and B to check for equal parity, while the double lines joining the spin measurement and $\sigma_z$ indicates an operation conditioned on measuring $|1\rangle$.}
\end{figure}

We now demonstrate how the parity measurement can be used to distil high-quality entanglement from two Bell pairs shared by two parties Alice (A) and Bob (B). Figure \ref{fig:purification}(a) shows a schematic of the purification process where two Bell pairs in a mixed state dominated by $|\phi^+\rangle_{13} | \phi^+ \rangle_{24}$, produce a single Bell pair with higher fidelity. The circuit for the purification process is shown in Fig.~\ref{fig:purification}(b) and proceeds as follows:
\begin{enumerate}
\item Alice performs the unitary transformation $U_A$ that maps the qubits $|0 \rangle \rightarrow (| 0 \rangle - i | 1 \rangle)/\sqrt{2}$ and $|1 \rangle \rightarrow (i | 0 \rangle + | 1 \rangle)/\sqrt{2}$, while Bob performs $U_B$ mapping $|0 \rangle \rightarrow (| 0 \rangle + i | 1 \rangle)/\sqrt{2}$ and $|1 \rangle \rightarrow (i | 0 \rangle + | 1 \rangle)/\sqrt{2}$.
\item Alice and Bob perform local parity measurements.
\item Alice and Bob compare their results. If they obtain the same parity they keep their qubits otherwise they discard both.
\item A $\pi/2$ rotation ($| 0 \rangle \rightarrow (| 0 \rangle + | 1 \rangle)/\sqrt{2}$ and $| 1 \rangle \rightarrow (-| 0 \rangle + | 1 \rangle)/\sqrt{2}$) is performed on qubits 2 and 4.
\item Alice and Bob measure qubits 2 and 4 and perform a $\sigma_z$ operation on qubits 1 and 3 respectively if they measure their qubit in the state $|1 \rangle$.
\end{enumerate}
The effect of this process on the density matrix of the qubit pair is the same as that reported in \cite{Deutsch1996PRL}, and leads to the final Bell pair between qubits 1 and 3 having higher fidelity if successful.

\section{Bell state analysis}

\begin{figure}[!th]
\includegraphics[width=0.7\columnwidth]{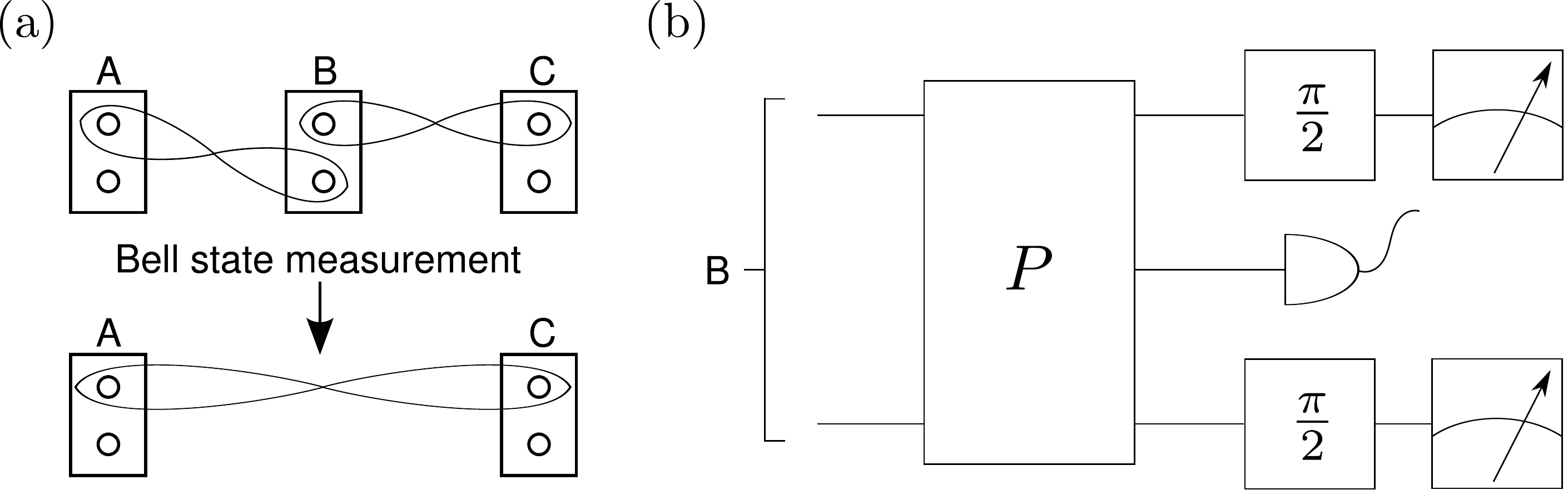}
\caption{\label{fig:BSA} (a) Schematic for entanglement swapping using photonic nodes. A Bell state measurement on the pair of qubits at B swaps the entanglement to form a Bell pair shared across A and C. (b) Circuit for Bell state analysis at B.}
\end{figure}

The parity measurement can also be used for Bell state measurements which can be used to swap entanglement between distant qubits in a quantum repeater network. Figure \ref{fig:BSA}(a) shows three nodes represented by boxes labelled A, B and C. Each of these boxes represents the MZI with two enclosed qubits (represented by circles) as shown in Fig.~\ref{fig:MZIparity}(a). Initially, one of the two qubits in node B is entangled with a qubit from node A and the other with a qubit from node C. A local Bell state measurement at node B leads to  qubits in A and C being entangled. The circuit to carry out this measurement is shown in Fig.~\ref{fig:BSA}(b). The parity measurement and rotation lead to the following Bell state mapping
\begin{equation}
\begin{split}
| \psi^+ \rangle &\rightarrow -| \phi^- \rangle | 1 \rangle_\textrm{ph} \\
| \psi^- \rangle &\rightarrow | \psi^- \rangle | 1 \rangle_\textrm{ph}\\
| \phi^+ \rangle &\rightarrow | \phi^+ \rangle | 0 \rangle_\textrm{ph}\\
| \phi^- \rangle &\rightarrow | \psi^+ \rangle | 0 \rangle_\textrm{ph}.
\end{split}
\end{equation}
From this expression it can be seen that the detection of a photon in state $|0 \rangle_{\rm ph}$ or $|1 \rangle_{\rm ph}$ differentiates between the $\psi$ and $\phi$ subspaces. A subsequent spin measurement of both qubits determines whether the Bell state was in the $+$ (qubits in different spin states) or $-$ (qubits in the same spin states) subspace. In total, these measurements thus project the qubits at node B into the Bell basis. As a result, the entanglement is swapped to nodes A and C.

\section{CZ gate}

We now show that a teleportation-based CZ gate can be realized using the parity measurement and classical communication channels. We start with the configuration shown in Fig.~3 of the main text and proceed with the details of the evolution. The logical qubits 2 and 4 are initially in an arbitrary pure state and the resource qubits 1 and 3 are in the Bell state $| \phi^+ \rangle_{13}$,
\begin{equation}
| \psi \rangle = | \phi^+ \rangle_{13} \left(c_{00} | 00 \rangle_{24} + c_{01} | 01 \rangle_{24} + c_{10} | 10 \rangle_{24} + c_{11} | 1 1 \rangle_{24} \right).
\end{equation}
Reordering the qubits to write them in terms of the local states at A and B gives
\begin{equation}
\begin{split}
| \psi \rangle = &c_{00} (|00 \rangle_{12}|00 \rangle_{34} + |10 \rangle_{12}|10 \rangle_{34}) + c_{01} (|00 \rangle_{12}|01 \rangle_{34} + |10 \rangle_{12}|11 \rangle_{34})\\
 + &c_{10} (|01 \rangle_{12}|00 \rangle_{34} + |11 \rangle_{12}|10 \rangle_{34}) + c_{11} (|01 \rangle_{12}|01 \rangle_{34} + |11 \rangle_{12}|11 \rangle_{34}).
\end{split}
\end{equation}
Performing a parity measurement on qubits 1 and 2 results in
\begin{equation}
\begin{split}
\textrm{Click on Detector 0 } \rightarrow \, &c_{00}|0 0 \rangle | 0 0 \rangle + c_{01}|0 0 \rangle | 0 1 \rangle c_{10}|1 1 \rangle + | 1 0 \rangle + c_{11}|1 1 \rangle | 1 1 \rangle,\\
\textrm{Click on Detector 1 } \rightarrow \,  &c_{00}|1 0 \rangle | 1 0 \rangle + c_{01}|1 0 \rangle | 1 1 \rangle + c_{10}|0 1 \rangle | 0 0 \rangle + c_{11}|0 1 \rangle | 0 1 \rangle,
\end{split}
\end{equation}
where for brevity we have dropped the state subscripts. Conditioned upon a click on Detector 1, we perform a $\sigma_x$ operation on qubits 1 and 3 which then leads to the same state as for clicks on Detector 0. This is followed by a Hadamard rotation on qubit 3 giving
\begin{equation}
\frac{1}{\sqrt{2}} \left[ c_{00} (| 0 0 \rangle | 0 0 \rangle + | 0 0 \rangle | 1 0 \rangle) + c_{01} (| 0 0 \rangle | 1 1 \rangle + | 0 0 \rangle | 0 1 \rangle) + c_{10} (| 1 1 \rangle | 0 0 \rangle - | 1 1 \rangle | 1 0 \rangle) + c_{11} (| 1 1 \rangle | 0 1 \rangle - | 1 1 \rangle | 1 1 \rangle) \right],
\end{equation}
which is then followed by a parity measurement on qubits 3 and 4
\begin{equation}
\begin{split}
\textrm{Clicks on Detector 0 } \rightarrow \,\, & c_{00} |0 0 \rangle | 0 0 \rangle + c_{01} | 0 0 \rangle | 1 1 \rangle + c_{10} | 1 1 \rangle | 0 0 \rangle - c_{1 1} | 1 1 \rangle | 1 1 \rangle \\
\textrm{Clicks on Detector 1 } \rightarrow \,\,  &c_{00} |0 0 \rangle | 1 0 \rangle + c_{01} | 0 0 \rangle | 0 1 \rangle - c_{10} | 1 1 \rangle | 1 0 \rangle + c_{1 1} | 1 1 \rangle | 0 1 \rangle ,
\end{split}
\end{equation}
Conditioned upon a click on Detector 1, we perform a $\sigma_x$ operation on qubit 3 and a $\sigma_z$ operation on qubit 2, which leads to the state being the same as that for a click on Detector 0. From the output state it can be seen that this results in a CZ gate if we erase qubits 1 and 3. We do this by performing Hadamard rotations and projectively measuring each qubit. This is followed by a $\sigma_z$ operation on qubits 2 and 4 on the condition of measuring $|1\rangle$. The Hadamard rotation and measurement on qubit 1 gives
\begin{equation}
\begin{split}
\textrm{Measure state }|0 \rangle \rightarrow \,\, & c_{00} | 0 \rangle_2 | 0 0 \rangle_{34} + c_{01} | 0 \rangle_2 | 1 1 \rangle_{34} + c_{10} | 1 \rangle_2 | 0 0 \rangle_{34} - c_{1 1} | 1 \rangle_2 | 1 1 \rangle_{34} \\
\textrm{Measure state }|1 \rangle \rightarrow \,\,  & c_{00} |0 \rangle_2 | 0 0 \rangle_{34} + c_{01} | 0 \rangle_2 | 1 1 \rangle_{34} - c_{10} | 1 \rangle_2 | 0 0 \rangle_{34} + c_{1 1} | 1 \rangle_2 | 1 1 \rangle_{34}.
\end{split}
\end{equation}
From here it can be seen that we get the same state in both cases if we perform a $\sigma_z$ operation on qubit 2 conditioned on measuring $| 1 \rangle$. Finally the Hadamard rotation followed by the measurement on qubit 3 gives
\begin{equation}
\begin{split}
\textrm{Measure state }|0 \rangle \rightarrow \,\, & c_{00} | 0 \rangle_2 | 0  \rangle_4 + c_{01} | 0 \rangle_2 |  1 \rangle_4 + c_{10} | 1 \rangle_2 |  0 \rangle_4 - c_{1 1} | 1 \rangle_2 |  1 \rangle_4 \\
\textrm{Measure state }|1 \rangle \rightarrow \,\, & c_{00} | 0 \rangle_2 | 0  \rangle_4 - c_{01} | 0 \rangle_2 |  1 \rangle_4 + c_{10} | 1 \rangle_2 |  0 \rangle_4 + c_{1 1} | 1 \rangle_2 |  1 \rangle_4,
\end{split}
\end{equation}
where it can again be seen that we get the same state if we perform a $\sigma_z$ operation on qubit 4 conditioned on measuring $| 1 \rangle$. This yields a CZ gate between qubits 2 and 4.

\section{CZ gate fidelity}

Our implementation of the CZ gate consists of two parity measurements on separate qubit pairs in addition to single qubit rotations. Each parity measurement can have two separate outcomes and therefore the CZ gate can proceed in four scenarios: measuring even/even, even/odd, odd/even and odd/odd parity. The fidelity of the CZ gate is therefore defined as
\begin{equation}
{\cal F}_\textrm{CZ} = \frac{P_{00} {\cal F}_{00} + P_{01} {\cal F}_{01} + P_{10} {\cal F}_{10} + P_{11} {\cal F}_{11}}{P_{00} +P_{01} +P_{10} +P_{11}},
\end{equation}
where $P_{ij}$ is the probability of measuring parity $i$ followed by parity $j$ and ${\cal F}_{ij}$ is the corresponding fidelity of the two measurements. Since the parity measurements occur on different subspaces, i.e. the first measurement is on qubits 1 and 2, while the second is on qubits 3 and 4, the probabilities and fidelities are simply given by the products of the individual probabilities and fidelities $P_{i j} = P_i P_j$ and ${\cal F}_{i j} = {\cal F}_i {\cal F}_j$. This means that, for two equivalent parity measuring circuits, the fidelity of the CZ gate is simply the square of the parity measurement fidelity ${\cal F}_\textrm{CZ} = {\cal F}^2$, and is not affected by the eventual erasure of qubits 1 and 3.

\begin{figure}[!th]
\includegraphics[width=0.5\columnwidth]{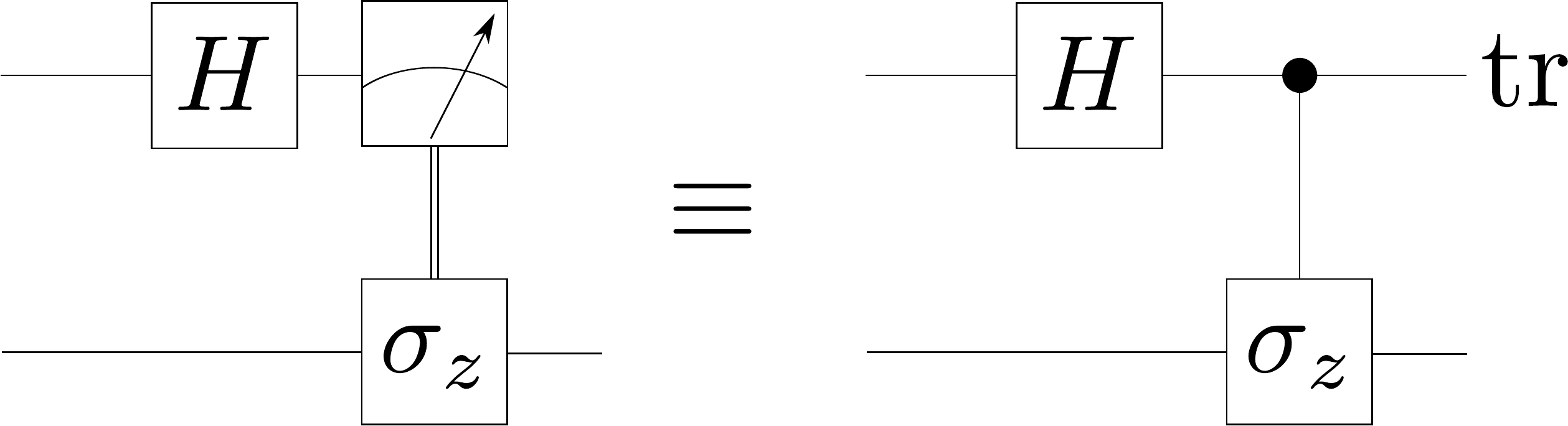}
\caption{\label{fig:eraseCZ} Since we assume single qubit operations to be ideal, erasing a qubit through a rotation, measurement, and a conditional rotation is equivalent to rotating it, performing an CZ operation, and tracing over the qubit.}
\end{figure}

We now show this explicitly for the term ${\cal F}_{10}$ using the Choi-Jamiolkowski input state and the two-click protocol. The process for the other terms is analogous. The Choi-Jamiolkowski input state is
\begin{equation}
\begin{split}
| \textrm{in} \rangle = \frac 1 2  | \phi^+ \rangle_{13} \left[ | \phi^+ \rangle_{24} | \phi^+ \rangle + | \phi^- \rangle_{24} | \phi^- \rangle + | \psi^+ \rangle_{24} | \psi^+ \rangle + | \psi^- \rangle_{24} | \psi^- \rangle \right],
\end{split}
\end{equation}
which we then rewrite in the reordered logical basis for qubits 1-4
\begin{equation}
\begin{split}
| \textrm{in} \rangle &= \frac 1 4 \left[ \left(| 0 0 0 0 \rangle_{1234} + | 0101 \rangle + |1010\rangle + | 1111 \rangle \right) | \phi^+ \rangle +  \left( | 0000 \rangle - | 0101 \rangle + | 1010 \rangle - | 1 1 1 1\rangle \right) | \phi^- \rangle \right]\\
&+ \frac 1 4 \left[ \left( | 0 0 0 0 1 \rangle + | 0100 \rangle  + | 1011 \rangle + | 111 0  \rangle \right) | \psi^+ \rangle +  \left( | 0 0 0 0 1 \rangle - | 0100 \rangle  + | 1011 \rangle - | 111 0  \rangle \right) | \psi^- \rangle \right],
\end{split}
\end{equation}
where for brevity we have only included the subscripts denoting the qubit order on the first term. We now perform a parity measurement using the two-click protocol on qubits 1 and 2 and consider obtaining clicks on Detector 1 after which we perform $\sigma_x$ rotations on qubits 1 and 3 as described in the previous section. Dropping the subscripts, we obtain the state
\begin{equation}
\begin{split}
\frac{\eta}{4 \sqrt{P_1}} \left[\left(| 1 1 1 1 \rangle + | 0 0 0 0 \rangle \right) | \phi^+ \rangle + (-| 1 1 1 1 \rangle + | 0 0 0 0 \rangle ) | \phi^- \rangle + (-| 1 1 1 0 \rangle + | 0 0 0 1 \rangle )| \psi^+ \rangle + (-| 1 1 1 0 \rangle + | 0 0 0 1 \rangle )| \psi^- \rangle \right],
\end{split}
\end{equation}
where we include detuning and loss errors, which according to (\ref{eq:out11}) can be introduced through $\eta = (1-t_0)(t_1-1)/4$ and $P_1 = |\eta|^2/2$. This is followed by a Hadamard rotation on qubit 3 and a parity measurement on qubits 3 and 4. Here we consider the parity measurement registering clicks on Detector 0. This leads to the state
\begin{equation}
\begin{split}
&\frac{\eta}{4 \sqrt{2} \sqrt{P_{1,0}}} \left[ \left( \epsilon | 11 01 \rangle - \alpha | 1111 \rangle + \alpha | 0 0 0 0 \rangle + \epsilon | 0 0 1 0 \rangle \right) | \phi^+ \rangle + \left( -\epsilon | 11 01 \rangle + \alpha | 1111 \rangle + \alpha | 0 0 0 0 \rangle + \epsilon | 0 0 1 0 \rangle \right) | \phi^- \rangle \right]\\
&+ \frac{\eta}{4 \sqrt{2} \sqrt{P_{1,0}}} \left[ \left( -\epsilon | 1110 \rangle + \alpha | 1100 \rangle + \alpha | 0 0 11 \rangle + \epsilon | 0 0 0 1 \rangle \right) | \psi^+ \rangle + \left( \epsilon | 11 10 \rangle - \alpha | 1100 \rangle + \alpha | 0 0 1 1 \rangle + \epsilon | 0 0 01 \rangle \right) | \phi^- \rangle \right],
\end{split}
\end{equation}
where $\alpha = (t_0 +t_1)/2$ and $\epsilon = (1+t_0)(1+t_1)/4$ and $P_{1,0} = |\eta|^2 (|a|^2 + |\epsilon|^2)/4$. As discussed in the previous section, we now erase qubits 1 and 3 and perform conditional $\sigma_z$ measurements on qubits 2 and 4. Although we can consider each measurement result individually it is simpler to construct a density matrix by replacing the measurement and conditional rotation with a gate and tracing over the qubits that are projected out. With ideal single qubit rotations, the two processes are equivalent as shown in Fig.~\ref{fig:eraseCZ}. For the first qubit, after performing the CZ gate, the state remains pure with the form
\begin{equation}
\begin{split}
&| \xi \rangle = \frac{\eta}{4 \sqrt{2}\sqrt{P_{1,0}}} \left[(\epsilon | 1 0 1 \rangle_{234} - \alpha | 1 1 1 \rangle + \alpha | 0 0 0 \rangle + \epsilon | 0 1 0 \rangle ) | \phi^+ \rangle + (-\epsilon | 1 0 1 \rangle_{234} + \alpha | 1 1 1 \rangle + \alpha | 0 0 0 \rangle + \epsilon | 0 1 0 \rangle ) | \phi^- \rangle \right]\\
&+ \frac{\eta}{4 \sqrt{2}\sqrt{P_{1,0}}} \left[(-\epsilon | 110 \rangle_{234} + \alpha | 1 00 \rangle + \alpha | 0 11 \rangle + \epsilon | 0 01 \rangle ) | \psi^+ \rangle + (\epsilon | 1 10 \rangle_{234} - \alpha | 1 00 \rangle + \alpha | 0 11 \rangle + \epsilon | 0 0 1 \rangle ) | \psi^- \rangle \right]
\end{split}
\end{equation}
Repeating this process on qubit 3 we obtain the density matrix $\rho_{10} = (| \xi_+ \rangle \langle \xi_+ | + | \xi_- \rangle \langle \xi_- |)/2$, where the subscripts $+$ and $-$ denote the states resulting from measuring $|0\rangle_3$ or $|1\rangle_3$ respectively, where
\begin{equation}
| \xi_{\pm} \rangle = \frac {\eta}{4 \sqrt{2 P_{1,0}}} \left\lbrace (\alpha \pm \epsilon) \left[ | 00 \rangle_{24} (|\phi^+ \rangle + | \phi^- \rangle) + |0 1 \rangle (|\psi^+ \rangle + | \psi^- \rangle )   \right] + (\alpha \mp \epsilon) \left[ | 11 \rangle (|\phi^- \rangle - | \phi^+ \rangle) + |10 \rangle (|\psi^+ \rangle - | \psi^- \rangle )  \right]  \right\rbrace.
\end{equation}
From this we can compute the CJ fidelity of the output density matrix using the ideal state $| \textrm{ideal} \rangle = (| \phi^- \rangle | \phi^+ \rangle + | \phi^+ \rangle | \phi^- \rangle + | \psi^+ \rangle | \psi^+ \rangle + | \psi^- \rangle | \psi^- \rangle)/2$. The $\epsilon$ terms cancel and the fidelity is
\begin{equation}
{\cal F}_{10} = \frac{|1-t_0|^2 |1-t_1|^2 |t_0+t_1|^2}{16 P_{1,0}}
\end{equation}
Using the definition of $P_{1,0}$ this becomes ${\cal F}_{10} = |t_0+t_1|^2/(|t_0+t_1|^2 + |1+t_0|^2|1+t_1|^2/4) = {\cal F}_1 {\cal F}_0$, where ${\cal F}_1=1$, and ${\cal F}_0$ is given by (\ref{eq:Fclick00}). Showing similar results for the other three parity combinations follows in a similar manner and leads to ${\cal F}_{\rm CZ} = {\cal F}^2$.

\end{document}